\documentclass{mn2e}
\input psfig.sty

\newcommand{\sax}{{\it Beppo\-SAX}}

\newcommand{\msun}{{\rm M}_{\sun}}
\newcommand{\xte}{{\it RXTE}}

\newcommand{\gro}{{\it CGRO}}

\title[Flares from Cygnus X-1]{Discovery of powerful millisecond flares from Cygnus X-1}

\author[Gierli\'nski \& Zdziarski]
{Marek Gierli\'nski$^{1,2}$ and Andrzej A. Zdziarski$^{3}$\\
$^1$Department of Physics, University of Durham, Durham DH1~3LE, UK;
Marek.Gierlinski@durham.ac.uk\\
$^{2}$Astronomical Observatory, Jagiellonian University, Orla 171, 30-244 Krak{\'o}w, Poland\\
$^3$Centrum Astronomiczne im.\ M. Kopernika, Bartycka 18,
00-716 Warszawa, Poland; aaz@camk.edu.pl\\}

\date{Accepted 2003 June 9. Received 2003 April 1.}

\pagerange{\pageref{firstpage}--\pageref{lastpage}}
\pubyear{2003}

\begin{document}

\maketitle

\label{firstpage}

\begin{abstract} We have found a large number of very strong flares in the 
available \xte/PCA data of Cyg X-1  (also seen in available HEXTE and BATSE 
data) with 13 flares satisfying our chosen threshold criterion, occuring both in 
the hard and the soft states. We analyze here in detail two of them. The 
strongest one took place in the soft state, with the 3--30 keV energy flux 
increasing 30 times with respect to the preceding 16-s average. The e-folding 
time is $\sim 7$ ms for the main flare and $\sim 1$ ms for its precursor. The 
spectrum strongly hardens during the flare. On the other hand, flares in the 
hard state have generally lower amplitudes and longer e-folding times, and their 
spectra soften during the flare, with the hardness of the spectrum at the flare 
peak similar for both types of the flares. The presence of the flares shows 
unusually dramatic events taking place in the accretion flow of Cyg X-1. On the 
other hand, the rate of occurence of hard-state flares shows they may represent 
a high-flux end of the distribution of shots present in usual lightcurves of Cyg 
X-1.  \end{abstract}


\section{Introduction}
\label{intro}

Among Galactic black-hole X-ray sources, Cyg X-1 is both the best-studied and 
the first one discovered (Bowyer et al.\ 1965). Its X-ray variability on various 
time scales has been considered to be relatively modest. In its dominant hard 
state, lightcurves for a wide range of time scales consist of repeating flares 
and dips, in which the flux departs from its average value by a factor of at 
most a few (Negoro, Miyamoto \& Kitamoto 1994; Feng, Li \& Chen 1999, hereafter 
F99; Zdziarski et al.\ 2002, hereafter Z02). The corresponding rms variability 
over the range of time scales of $\la 10^3$ s is typically $\la 50$ per cent 
(Lin et al.\ 2000). In the soft state, the overall X-ray variability corresponds 
to an even lower rms, and it can be decomposed into a constant blackbody-like 
component and a tail varying within a factor of $\sim 2$ (Churazov et al.\ 2001; 
Z02). Recently, occasional departures from the above patterns have been reported 
by Stern, Beloborodov \& Poutanen (2001) and Golenetskii et al.\ (2003), who 
reported several long outbursts lasting $\ga 10^3$ s during which the 15--300 
keV flux increased about an order of magnitude above the average.

On short timescales, (relatively weak) flares or shots are seen in the X-ray 
light curves, although the description of the variability of Cyg X-1 in terms of 
this model is not unique (e.g., Lochner, Swank \& Szymkowiak 1991; Negoro et 
al.\ 1994, 1995; Pottschmidt et al.\ 1998; F99) as well as the shots cannot 
correspond to independent, randomly occurring, events (e.g., Uttley \& McHardy 
2001). The shot profiles depend on the spectral state, with shots in the soft 
state being faster than those in the hard state (F99).

In this {\it Letter}, we report the discovery with the \xte/PCA of very strong 
subsecond flares occuring both during the hard and the soft states (with the 
count rate increasing by a factor up to $\sim 20$). The strongest and fastest 
flare occured during the extended soft state in 2002 (Z02). During that flare, 
the 3--30 keV flux increased by a factor of $\sim 30$. The flare was preceded by 
a weaker precursor, which showed an e-folding time scale of $\la 2$ ms. This is 
similar to the light travel time across an inner accretion disc around a
$10\msun$ black hole (1 ms $\simeq 20 GM/c^3$), and it is $\la 1/2$ of the 
Keplerian period on the minimum stable orbit in the Schwarzschild metric. It 
also closely corresponds to the e-folding time of the 1999 flare of Sgr A$^*$ 
(Baganoff et al.\ 2001) of $\sim 400$ s, which is $\sim 30 GM/c^3$ for a 
$3\times 10^6\msun$ black hole. Although variability on ms time scales in Cyg 
X-1 was reported from early experiments (Rothschild et al.\ 1974, 1977; Boldt 
1977; Meekins et al.\ 1984), its reality was questioned (Press \& Schechter 
1974; Weisskopf \& Sutherland 1978; Chaput et al.\ 2000). Also, no power on time 
scales $<2$ ms has been detected in the PCA data for Cyg X-1 (Revnivtsev, 
Gilfanov \& Churazov 2000), which appears to be generally the case for 
black-hole binaries (Sunyaev \& Revnivtsev 2000).  

The occurence of strong flares during the soft state is especially interesting 
because that state of black-hole binaries is likely to be the stellar 
counterpart of Narrow Line Seyfert 1s (NLS1s, Pounds, Done \& Osborne  1995). 
This correspondence has sometimes been questioned on the ground of the 
relatively weak variability in the soft state of Cyg X-1.  However, the 
variability of the soft-state flare reported here does resemble, in fact, the 
extreme variability of NLS1s (e.g., Boller et al.\ 1997; Brandt et al.\ 1999).

\section{Data Analysis}
\label{data}

We have searched the entire \xte/PCA Cyg X-1 public data (2.3 Ms available as of 
2003 March) for fast flares using the following algorithm. We extract the PCA 
Standard-1 lightcurves (from all layers and detectors available) with 0.125 s 
timing resolution in 128~s segments. For each segment, we calculate the average 
count rate, $\langle C \rangle$, and its standard deviation, $\sigma$ (which is 
dominated by the intrinsic source variability, i.e., $\gg$ the Poissonian 
$\sigma$ of the count rate). Then, we look for events with excess count rate 
$>10\sigma$ over the average in each segment. This condition is chosen to 
correspond to events much stronger than those expected from a normal 
distribution of fluctuations of any nature [the relevant value of the errror 
function is ${\rm erfc}(10)\simeq 2\times 10^{-45}$]. Table 1 gives a log of the 
13 fast strong flares we have found.

%=============

\begin{table} 
\label{tab:flares}
\centering 
\begin{minipage}{80mm}

\caption{Strong flares in Cyg X-1 from {\it RXTE}. The last 
colum gives the maximum excess flux divided by the standard 
deviation during 128 s segment of a 0.125-s lightcurve containing the 
flare. We list all flares where it is $>10$.}
\begin{tabular}{rccc}
\hline
No. & Obs.\ ID & UT$_{\rm peak}$ & $C_{\rm peak} - \langle C\rangle 
\over \sigma$ \\
\hline
1 & 10236-01-01-00 & 1996-12-16 06:55:19.8 & 11.5 \\
2 & 10236-01-01-00 & 1996-12-16 13:10:31.3 & 10.1 \\
3 & 10236-01-01-020 & 1996-12-16 16:31:55.8 & 11.8 \\
4 & 10236-01-01-020 & 1996-12-16 22:56:23.4 & 10.1 \\
5 & 20173-01-01-00 & 1997-01-17 00:51:35.3 & 11.4 \\
6 & 20173-01-01-00 & 1997-01-17 05:46:21.3 & 10.2 \\
7 & 30157-01-04-00 & 1997-12-30 18:48:41.0 & 11.2 \\
8 & 30157-01-06-00 & 1998-01-15 23:16:55.3 & 10.5 \\
9 & 30157-01-10-00 & 1998-02-14 00:14:58.8 & 11.6 \\
10 & 40100-01-12-03 & 1999-11-24 03:01:09.6 & 10.7 \\
11 & 40100-01-19-00 & 1999-12-30 22:41:31.3 & 10.2 \\
12 & 40102-01-01-030 & 2000-01-09 01:15:48.5 & 10.3 \\
13 & 70414-01-01-02 & 2002-07-31 00:06:30.5 & 12.8 \\
\hline
\end{tabular}
\end{minipage}
\end{table}

%============= FIG 1

\begin{figure}
\centerline{\psfig{file=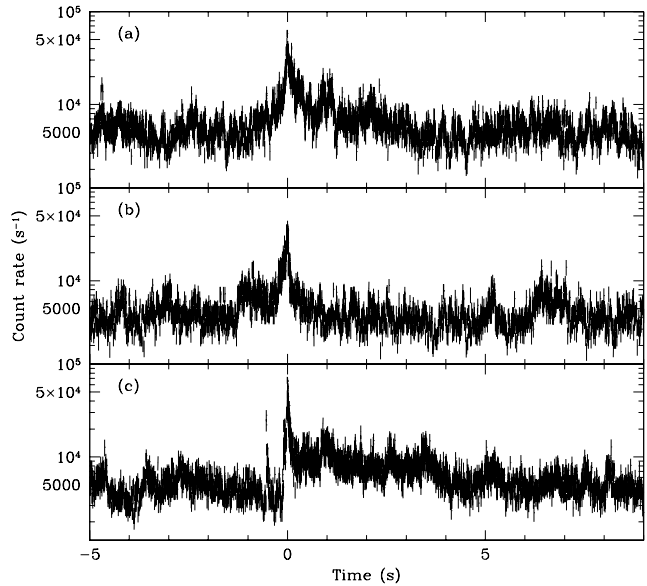,width=8.5cm}}
\caption{The PCA 2--60 keV lightcurves of the flares 1, 9 and 13 of Table 
1. The first two (with 5 PCUs on) and the third (with 4 PCUs on) occured during the hard and soft state, respectively. Each point has an accuracy better than 20 
per cent, which required rebinning of the original 2-ms lightcurve.}
\label{3flares}
\end{figure}

%=============

Since the bin length of 0.125 s is rather long in comparison to the ms 
time scales present in the flares, the peak intensities of some short flares may 
be significantly diminished by averaging over the bin duration. Therefore, the 
procedure we apply may not find all short events above a given threshold. 
Thus, the number of flares given here represents a lower limit on the 
actual $>10\sigma$ population.

For in-depth analysis, we have selected the flares 1 and 13, in the hard and 
soft state (see below), respectively. For both, we have extracted the PCA 
lightcurves from Single-Bit and Event modes in three energy bands (2--5.1, 
5.1--13, 13--60 keV for flare 1, and 2--5.7, 5.7--14.8, 14.8--60 keV for flare 
13). We dynamically adjust the bin length in order to limit the statistical 
errors. The lightcurves were corrected for the dead time of the detectors using 
a standard prescription given by the \xte\/ team (both VLE and non-VLE and 
taking into account the loss of a Propane layer in PCU0 after 2000 May 13). This 
correction reaches 23 per cent at the peak of the flare 13, and it is less for 
other flares. We assumed the dead-time effects to be energy independent. A 
constant background level estimated from the Standard 2 data has been 
subtracted. The background rate is $\ll$ that of the source even outside the 
flares. We note that the events were detected by all the PCA detectors available 
as well as by the HEXTE, which excludes their origin from high-energy particles 
hitting a detector. The HEXTE data have also been corrected for the background and dead time. We plot their lightcurves in Fig.\ \ref{3flares} together 
with that of flare 9 shown for comparison. For the flares 1 and 3, we have also been able to obtain simultaneous 1-s resolution BATSE data (B. Stern, private communication). The BATSE lightcurves in the channel $<50$ keV show increases of the countrate in the 1-s interval containing the flare with respect to the     preceeding one at the level of $4.8\sigma$, $5.6\sigma$, respectively. This further confirms the reality of the flares. 

We calculate the hardness ratios between the above energy bands. Then, we use an 
absorbed power-law model in {\sc xspec} to simulate PCA spectra for different 
values of the spectral index, $\Gamma$. By comparing the count rate and hardness 
ratios in the lightcurves and in the simulated spectra, we estimate unabsorbed 
$\Gamma$ and the energy flux in each time bin. A power law with that $\Gamma$ 
yields then the same hardness ratio as the actual spectrum, thus representing 
the overall spectral hardness (see Z02). We also extract energy spectra within 
200 and 70 ms around the peak for flares 1 and 13, respectively, from Binned and 
Event modes of the PCA. Unfortunately, low energy channels in the Binned modes 
suffer from overflowed counters at high count rates, so only the data at $\ga$ 
10 keV are usable.

\begin{figure*}
\centerline{\psfig{file=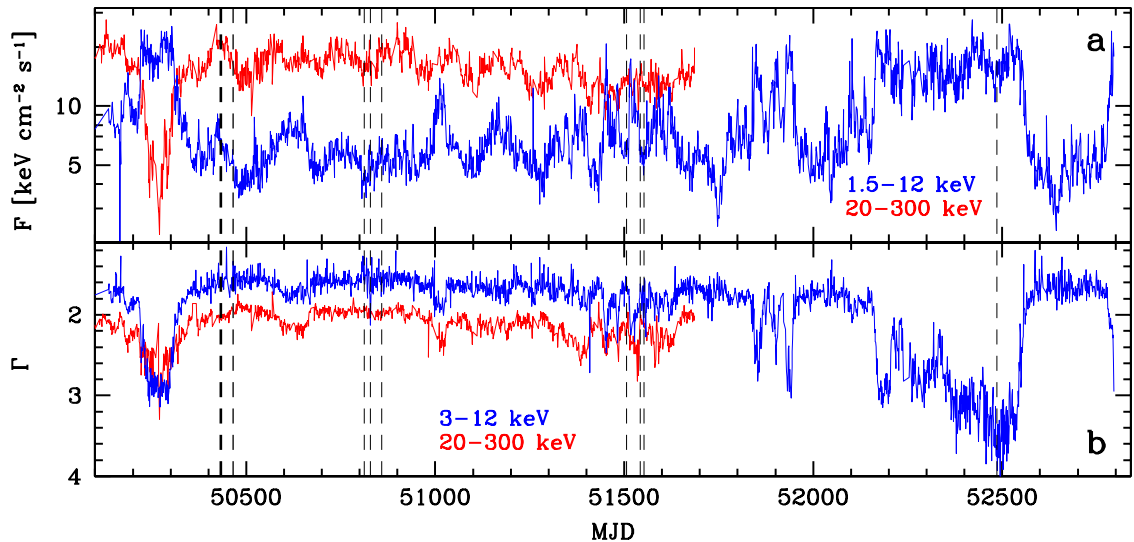,width=15.cm}}
\caption{(a) The ASM (blue, till 2003 June 5) and BATSE (red) lightcurves with the dashed lines showing the times of the occurence of the flares. The line width is proportional to the number of flares within a given day. (b) The corresponding average spectral indices.  Note the recent state transitions in 2001 September, 2002 October and 2003 May/June.}
\label{asm_batse}
\end{figure*}

Fig.\ \ref{asm_batse} shows ASM and BATSE lightcurves showing the times of the 
occurence of the flares. Based on the ASM data, 1.89 Ms, 0.15 Ms, and 0.21 Ms of 
the studied exposure time correspond to the hard, intermediate and soft state, 
respectively (see Z02 for the definitions). Notably, 12 flares occured in the 
rather average hard state, with the 3--12 keV and 20--300 keV hardness 
corresponding to $\Gamma\simeq 1.6$ and 2.0, respectively, and outside of any 
strong peaks of the ASM flux. The corresponding PCA data yield the intrinsic 
indices (corrected for absorption and reflection) in a narrow range around 
$\Gamma\simeq 1.7$. On the other hand, the last flare occured during a part of 
the extended 2002 soft state when the 3--12 keV slope was the softest ever, 
$\Gamma\ga 3.5$. 

\section{Flares in the soft state}
\label{soft}

\begin{figure}
\centerline{\psfig{file=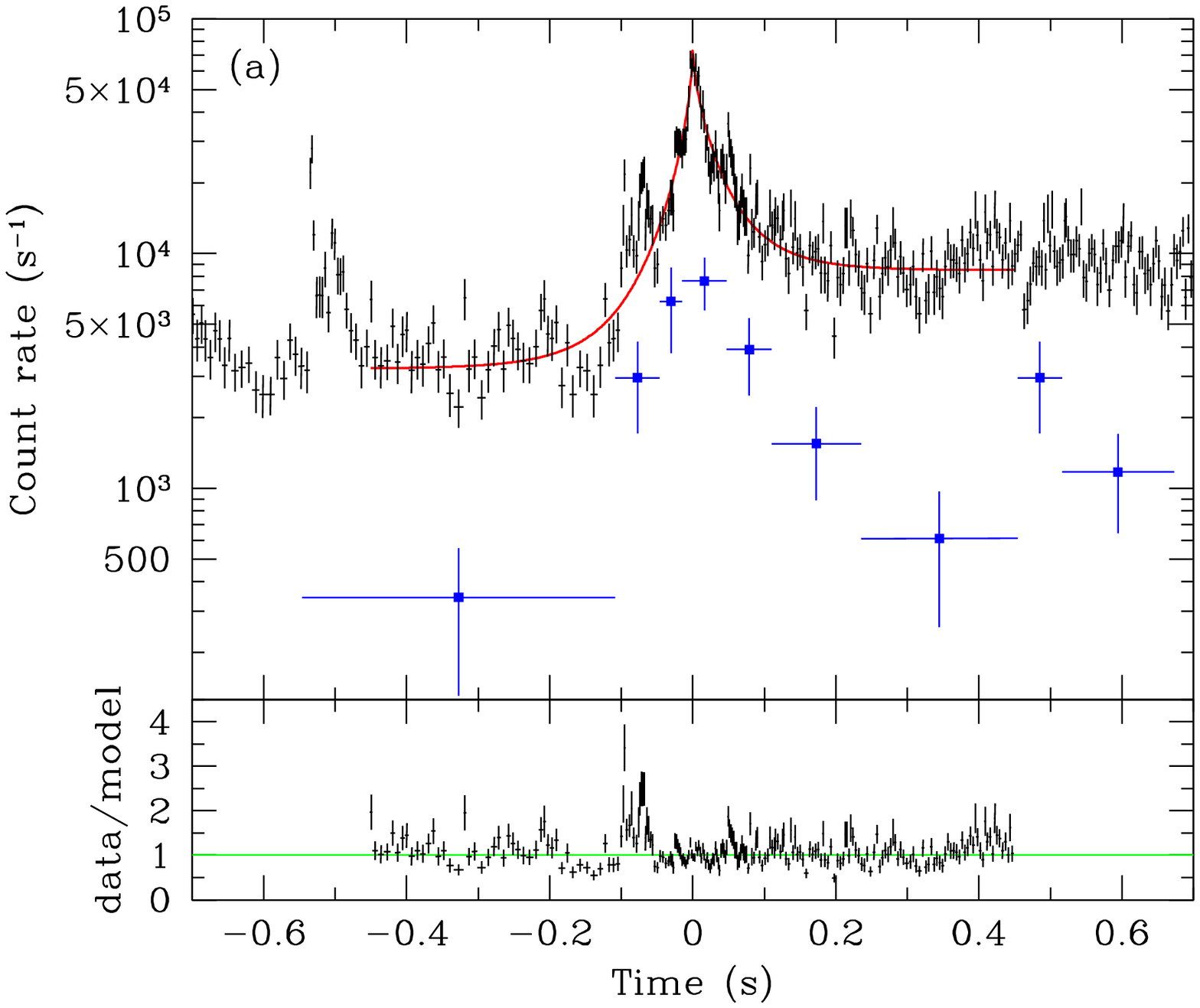,width=8.cm}}
\centerline{\psfig{file=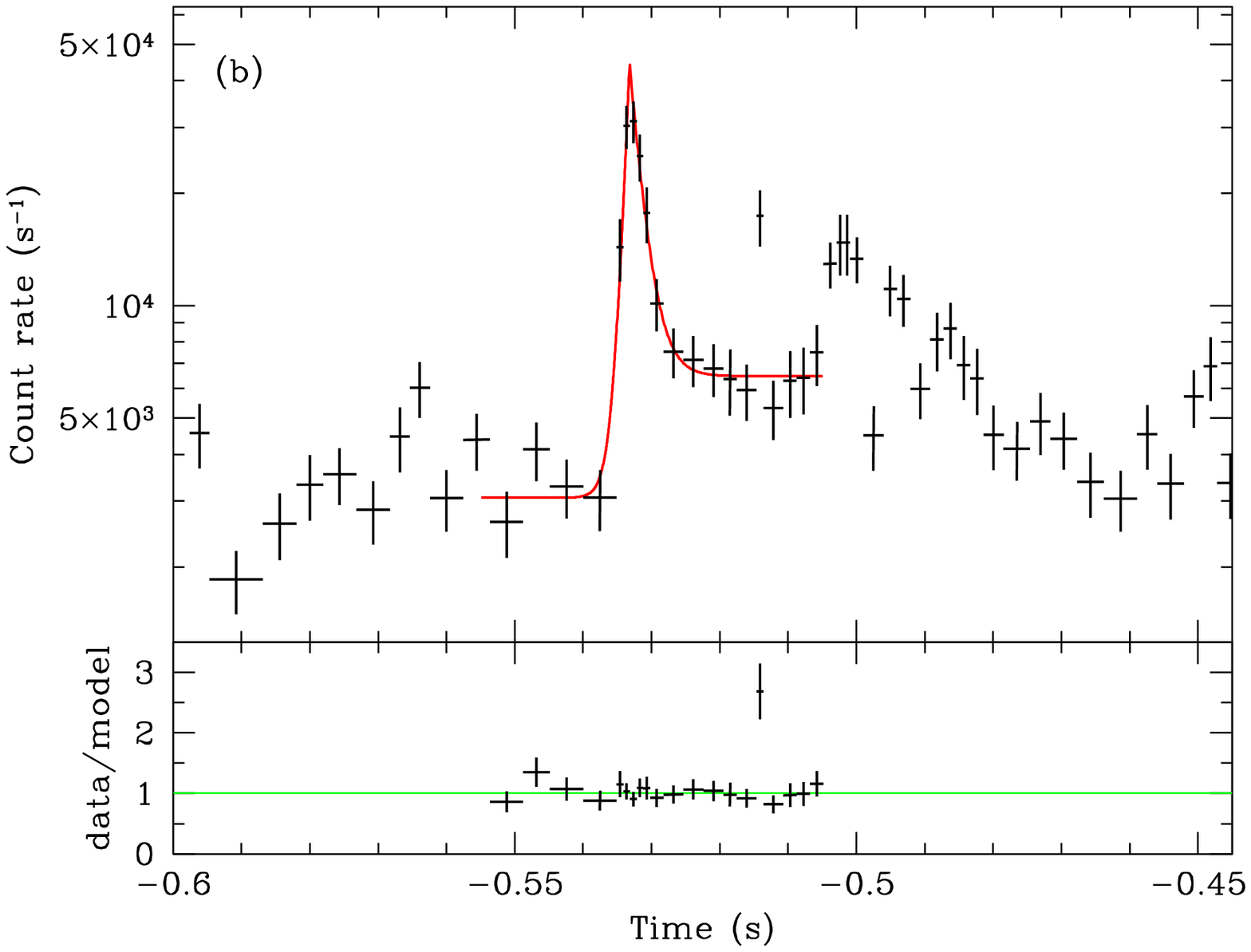,width=8.cm}}
\caption{(a) The 2--60 keV PCA and 15--60 keV HEXTE (error bars with filled squares, multiplied by a factor of 20) lightcurves, and PCA residuals of the flare 13 fitted (solid curve) by stretched exponential rise/decline. The average PCA count rate immediately preceding the flare was $\sim 3000$ s$^{-1}$, which first increased to $\sim 4\times 10^4$ s$^{-1}$ over $\sim 2$ ms during the precursor, and then to the peak of $\sim 7\times 10^4$ s$^{-1}$ with the FWHM $\simeq 25$ ms. The flare was then followed by a period of enhanced flux for about 10 s. (b) A detailed view of the profile of the precursor, fitted 
(the solid curve) by exponential rise/decline.}
\label{f13}
\end{figure}

The lightcurve of the flare 13 is shown in Figs.\ \ref{3flares}c and \ref{f13}. 
Fig.\ \ref{f13} gives detailed PCA profiles of both the main flare and the 
precursor, as well as the HEXTE lightcurve, very similar in shape to the PCA one. The peak of the count rate profile of the main flare is not well 
fitted by an exponential function, and therefore we used a stretched 
exponential, $C_0 \exp[-(|\Delta t|/\tau)^\beta]+ C_1$, where $\beta= 0.65\pm 
0.05$, $0.70\pm 0.10$ and $\tau=17\pm 5$, $21\pm 6$ ms before and after the 
peak, respectively. Even this function does not provide a good overall fit 
($\chi^2/\nu=590/220$), as the main flare appears to be superposition of several 
subflares much faster, $\ga 2$ ms, than the overall profile. If we fit the peak 
profile with the sum of two exponentials and a constant on each side of the peak 
(Negoro et al.\ 1994; F99), the shorter time scale is $\tau\simeq 7$ ms for both 
rise and decline, which is comparable to those obtained by F99 for the average 
soft-state shot (with the relative amplitude $\la 2$). The residuals for this 
model are very similar to those shown in Fig.\ \ref{f13}a. After the initial 
fast decline, we see a slow return to the average level (see Fig.\ 
\ref{3flares}c) with the e-folding time of $3.2\pm 0.6$ s. On the other hand, 
the precursor profile (Fig.\ \ref{f13}b) shows the rise and decline with the 
e-folding time as short as 1 ms and 2 ms, respectively. 

%=============

\begin{figure}
\centerline{\psfig{file=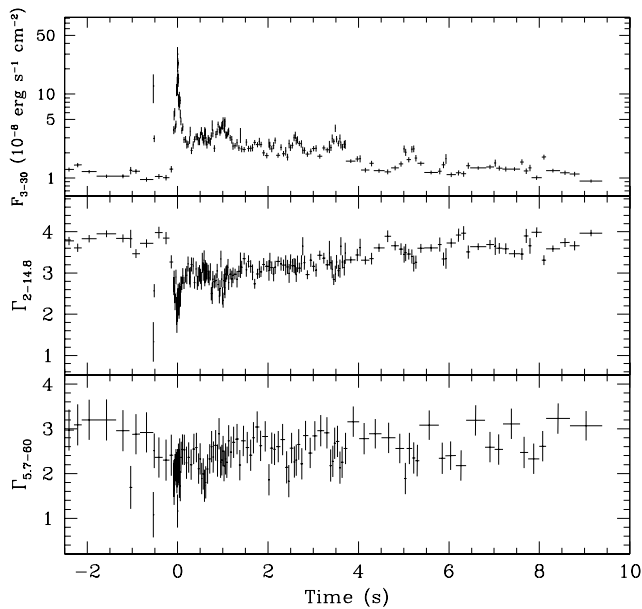,width=8.5cm}}
\caption{Evolution of the 3--30 keV flux and the average spectral 
indices in two bands for the flare 13.}
\label{f13_fg}
\end{figure}

The 3--30 keV flux during 2 ms containing the peak of the flare reached $(3.0\pm 
0.6)\times 10^{-7}$ erg cm$^{-2}$ s$^{-1}$, which is $\sim 30$ times the 
corresponding average flux in a 16 s period before the flare [and corresponds to 
$L\simeq (1.4\pm 0.3)\times 10^{38}$ erg s$^{-1}$ at $d=2$ kpc, which distance 
we assume hereafter]. Fig.\ \ref{f13_fg} shows the 3--30 keV energy flux profile 
as well as the average spectral indices in the spectral bands of 2--14.8 keV and 
5.7--60 keV. We see a strong hardening in both bands during both 
the main flare and the precursor, with $\Gamma_{2-14.8}$ decreasing from $\sim 
4$ to $2.2\pm 0.1$ during 10 ms around the peak of the flux, and then correlated 
with the flux during the $\sim 10$ s-long decline. The evolution of 
$\Gamma_{5.7-60}$ is from $\sim 3$ to $1.7\pm 0.2$ at the peak 10 ms. We note 
that an even stronger hardening occured during the precursor. All these changes 
are much stronger than a moderate hardening found by F99 for the soft-state 
shots (with the amplitude $\la 2$).

%=============

\begin{figure}
\centerline{\psfig{file=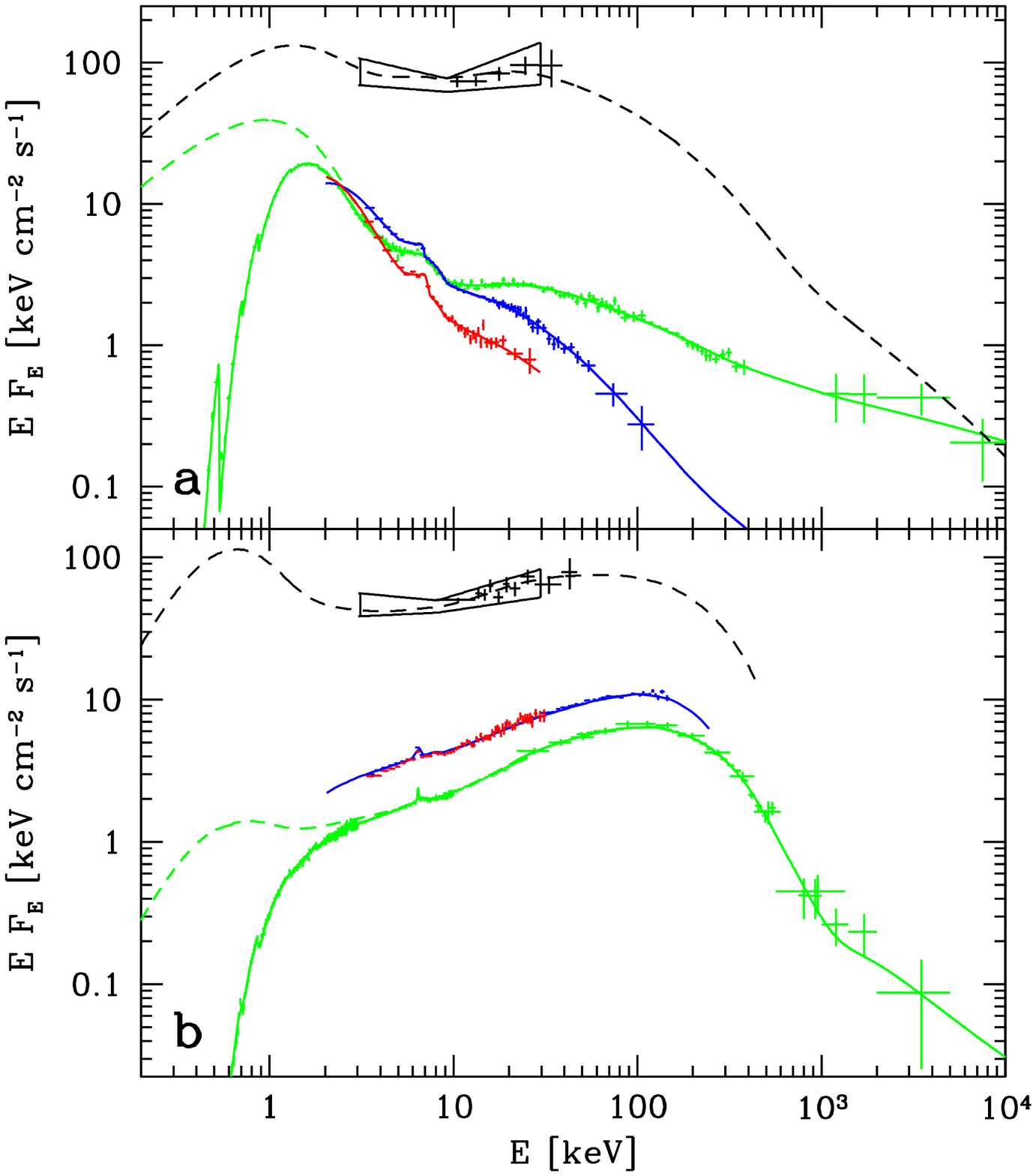,width=8.5cm}}
\centerline{\psfig{file=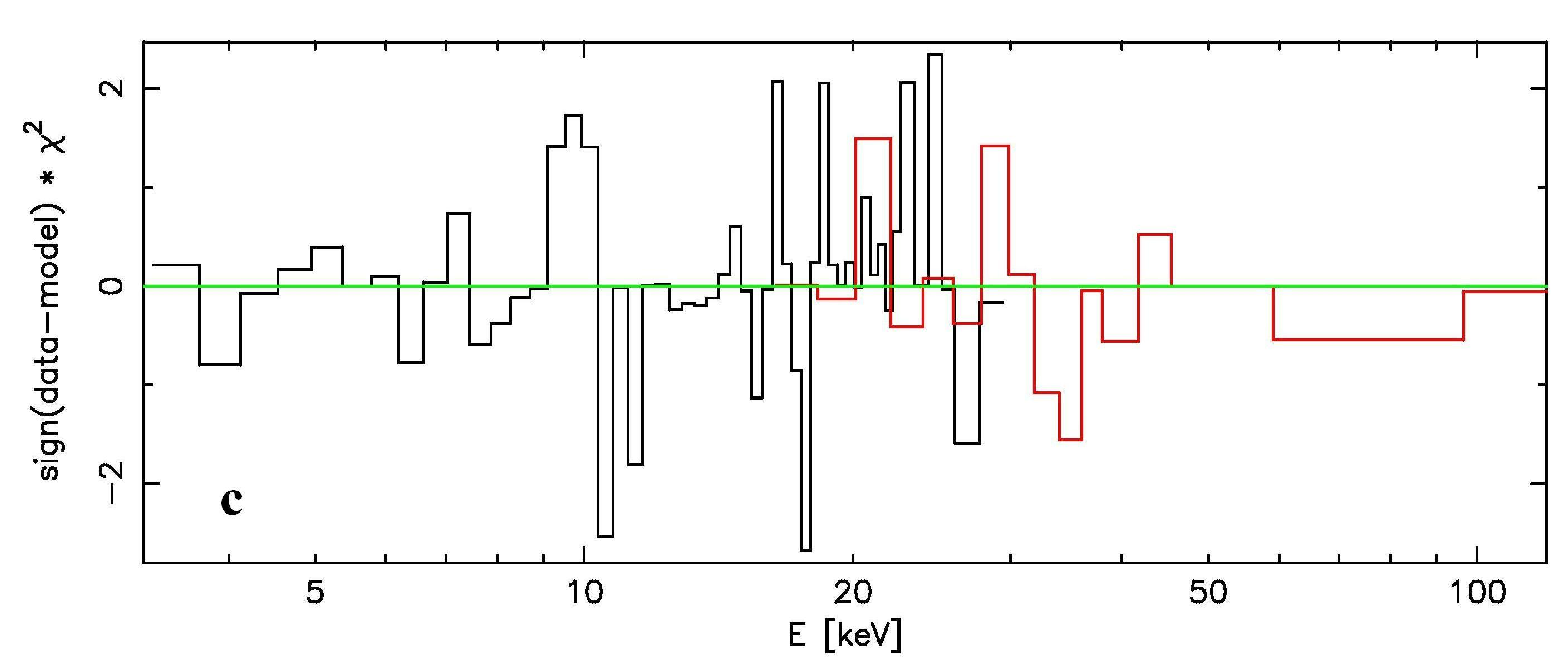,width=6cm}} 
\caption{Spectra related to the flare (a) on 2002-07-31, with the 1996 soft state spectrum from \sax\/ and \gro\/ shown for comparison in green; and (b) on 1996-12-16, with the average hard-state spectrum from \gro\/ and a \sax\/ spectrum matching the \gro\/ data at $\la 20$ keV shown in green. On both panels, the green dashed curve shows the corresponding intrinsic spectrum before 
absorption. The PCA/HEXTE average spectra during the 
observations containing the flares, and the PCA spectra from 16 
s before the flare are shown in blue and red, respectively. The 
black error contours show the spectra at the peak of the flare 
estimated from the ratios of count rates in three energy bands, 
and the black crosses give the PCA spectra from $\sim 0.1$--0.2 
s around the peak of the flare and renormalized to the peak 
flux. The black dashed curves show possible blackbody/Compton 
scattering models. (c) Residuals to the hybrid-Compton model for the 2002 July 30--31 observation.}
\label{spectra}
\end{figure}

Given the values of the flux and spectral indices at the peak, we can plot 
schematically the peak 3--30 keV spectrum in Fig.\ \ref{spectra}a. We also show 
the PCA spectrum from a 70 ms period containing the flare, which we plot 
normalized to the peak flux. Fig.\ \ref{spectra}a also shows the PCA spectrum 
from a 16 s interval preceding the flare as well as the average PCA/HEXTE 
spectrum from the entire observation of 2002 July 30--31. For comparison, we 
also show the broad-band spectrum of the soft state in 1996 June from \sax\/ 
and \gro\/ (McConnell et al.\ 2002). 

%=============

The last three soft-state spectra have been fitted with a hybrid Comptonization 
model, in which disc blackbody emission is upscattered by electrons with a 
Maxwellian distribution and a non-thermal tail (Coppi 1999; Gierli\'nski et al.\ 
1999). The tail is formed due to electron acceleration at a power-law rate with 
an acceleration index, $\Gamma_{\rm acc}$. Then, the steady-state electron 
distribution is solved for self-consistently, taking into account all important 
processes. The 2002 soft-state spectra are adequately described by this model. 
In particular, $\chi^2/\nu=75/81$ (assuming a 1 per cent systematic error for 
the PCA) for the fit to the PCA/HEXTE spectrum from the entire 2002 July 31 
observation, with low $\chi^2$ residuals shown in Fig.\ \ref{spectra}c. We have 
found that the 2002 spectra have much softer tails, corresponding to 
$\Gamma_{\rm acc}\simeq 3.9\pm 0.6$, compared to $\Gamma_{\rm acc}\sim 2.5$ in 
1996 June (Gierli\'nski et al.\ 1999; Frontera et al.\ 2001; McConnell et al.\ 
2002). Also, the luminosity in the tail is significantly lower than that during 
the 1996 soft state. In both cases, that power is much less than that in the 
blackbody disc emission, shown for the 1996 spectrum (which was measured at soft 
X-rays) by the green dashed curve in Fig.\ \ref{spectra}a. 

The question arises what is the nature of the spectrum during the flare. We note 
that transient black-hole binaries go through the so-called very high state at 
high luminosities (e.g.\ Miyamoto et al.\ 1991), with a high amplitude of the 
high-energy tail with respect to the blackbody. Recently, Gierli\'nski \& Done 
(2003) have shown that the very high state spectrum of the transient XTE 
J1550--564 is well fitted by the hybrid model with the ratio between the power 
in the Comptonizing plasma to that in the blackbody disc of $\sim 1$. Then, the 
tail starts near the peak of the blackbody spectrum. Motivated by this, we have 
looked into the possibility that the flare spectrum is of similar nature, and 
found this indeed plausible. The black dashed curve in Fig.\ \ref{spectra}a 
shows the intrinsic spectrum of a possible model, with the blackbody component 
higher by a factor of $\sim 2$ than that during the 1996 soft state. The 
presence of non-thermal electrons is not constrained by the data, and we assumed 
that the power supplied to electron acceleration is the same as that in electron 
heating (similarly to the case of XTE J1550--564, Gierli\'nski \& Done 2003). 
The hardening at $\sim 10$ keV is due to the onset of Compton reflection 
(Magdziarz \& Zdziarski 1995), assumed here to correspond to a solid angle of 
$2\upi$. 

The bolometric flux of this model is $9\times 10^{-7}$ erg cm$^{-2}$ s$^{-1}$, 
which is $\sim 6$ times the bolometric flux of the intrinsic emission observed 
by \sax\/ and \gro\/ in 1996, and corresponds to $L\simeq 4.5\times 10^{38}$ erg 
s$^{-1}$, i.e.\ $\sim 0.3$ of the Eddington luminosity of a $10\msun$ star. The 
physical origin of the flare may be related to a disc instability close to its 
inner edge, in which there is a sudden conversion of energy accumulated in the 
Keplerian disc into magnetic heating of a hot plasma (e.g., Machida \& Matsumoto 
2003), or to flipping among multiple state of the accretion flow with large 
amplitude on short dynamical timesclaes as exhibited in recent MHD simulations 
by Proga \& Begelman (2003). Another possibility is accumulation of a very large 
amount of energy in a flare above the disc surface and its subsequent fast 
release (e.g., Beloborodov 1999).

Interestingly, we also found two weaker flares (with the excess 
flux divided by rms of $\sim 8$, thus not listed in Table 1) in the 2002 soft 
state, during which the X-ray spectrum softened. This is opposite to the case 
above, indicating more than one physical scenario leading to flaring in the soft 
state. 

\section{Flares in the hard state}
\label{hard}

\begin{figure}
\centerline{\psfig{file=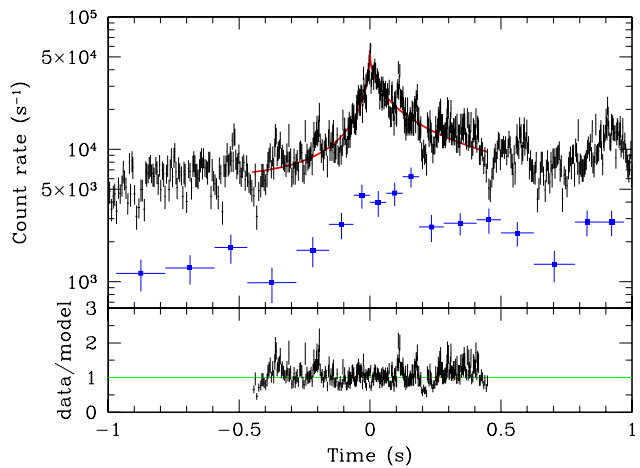,width=8.5cm}}
\caption{The 2--60 keV PCA and 15--60 keV HEXTE (error bars with filled squares, multiplied by a factor of 7) lightcurves, and PCA residuals of the flare 1 fitted (solid curve) by stretched exponential rise/decline. }
\label{f01}
\end{figure}

Figs.\ \ref{3flares}a and \ref{f01} show the count rate history of the first 
flare on 1996 Dec.\ 16. By fitting by a streched exponential, we obtain $\tau = 
27\pm 10$, $79\pm 24$ ms and $\beta = 0.50\pm 0.08$, $0.45\pm 0.10$ during the 
rise and decline, respectively. Although relatively strong, this flare is 
representative for other hard-state flares, see, e.g. the profile of flare 9 in 
Fig.\ \ref{3flares}b. Note also the appearance of four flares within 16 hr 
on 1996 Dec.\ 16, and two flares within 5 hr on 1997 Jan.\ 17. This indicates flare clustering with the correlation time of several hours.

%=============  FIG 7

%=============

\begin{figure}
\centerline{\psfig{file=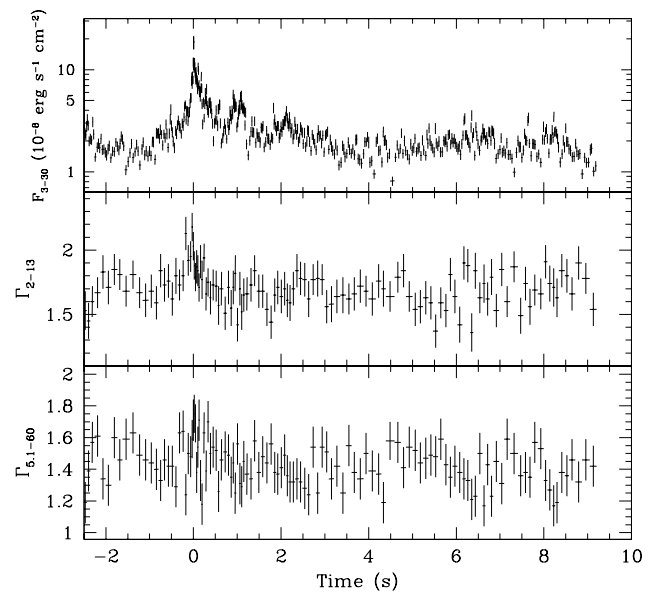,width=8.5cm}}
\caption{Evolution of the 3--30 keV flux and the average 
spectral indices in two spectral bands for the hard-state flare 
of 1996-12-16 (No.\ 1 in Table 1).}
\label{f01_fg}
\end{figure}

Fig.\ \ref{f01_fg} shows the corresponding profile  in energy flux units as well 
as the average spectral indices in the spectral bands of 2--13 keV and 5.1--60 
keV. The  3--30 keV flux during 2 ms containing the peak reached $(1.9\pm 
0.3)\times 10^{-7}$ erg cm$^{-2}$ s$^{-1}$, which is $11\pm 2$ times the 
corresponding average flux in a 16 s period before the flare, and corresponds to 
$L\simeq (9\pm 1)\times 10^{37}$ erg s$^{-1}$. We now see a softening during the 
flare (more pronounced in the 2--13 keV band), with $\Gamma_{2-13}$ and 
$\Gamma_{5.1-60}$ increasing from $\sim 1.7$ to $2.0\pm 0.1$ and from $\sim 1.5$ 
to $1.7\pm 0.1$, respectively, averaged over 33 ms containing the peak. A 
similar softening during the shots in the hard state has been found by Negoro et 
al.\ (1994) and F99. Note that the X-ray spectrum at the peak is rather similar 
to the corresponding one of the soft-state flare. 

%=============  FIG 8

%=============

Given the values of the flux and spectral indices at the peak, we can plot 
schematically the peak 3--30 keV spectrum in Fig.\ \ref{spectra}b, together with 
the PCA spectrum from a 16 s interval preceding the flare as well as the average 
PCA/HEXTE spectrum from the entire observation of 1996 Dec.\ 16. We also 
obtained the PCA spectrum from 0.2 s containing the flare peak, which is shown 
in Fig.\ \ref{spectra}b, and is fully compatible with the estimate from 
broad-band fluxes. For comparison, we also show the average hard-state spectrum 
from \gro\/ and \sax\/ (McConnell et al.\ 2002). Given the form of the flare 
spectra, a large increase of the flux of seed photons for Comptonization is 
required, as illustrated by the black dashed curve in Fig.\ \ref{spectra}b. This 
model is based on those for the hard state by Frontera et al.\ (2001) and Di 
Salvo et al.\ (2001), where there are two thermal Comptonization regions, 
dominating in soft and hard X-rays, respectively. The bolometric flux of this 
model is $\sim 8\times 10^{-7}$ erg cm$^{-2}$ s$^{-1}$, which is $\sim 20$ times 
the average bolometric flux in the hard state (Z02), and corresponds to $L\simeq 
3.7\times 10^{38}$ erg s$^{-1}$. Note that the similarity of the peak spectrum 
of this flare to that in the soft state implies that a hybrid model similar to 
that shown in Fig.\ \ref{spectra}a is also possible. Notably, the $\ga 10^3$-s 
flares found by Goleneetski et al.\ (2003) also show similar peak spectra and 
fluxes. 

\section{Statistics of the flares}
\label{shots}

An important question is that of the relation of our flares to the much weaker 
peaks seen in X-ray lightcurves of Cyg X-1. Apart from the issue of the 
applicability of the shot-noise model to Cyg X-1, we can still consider the 
distribution of the relative amplitudes of those peaks, or shots, with respect 
to the local average. That was studied by Negoro et al.\ (1995) and Negoro \& 
Mineshige (2002) for the hard state. The overall shape and time scales of their 
shots, although with much lower amplitudes, is relatively similar to that of the 
12 flares found here. Those authors have found good fits to the distribution of 
the peak shot count rates in 31.25 ms time bins by either an exponential or a 
log-normal function (with a power law strongly ruled out). Their results for 
$1.5\la r\la 3.5$, where $r$ is the ratio of the shot peak count rate to the 
average, can be expressed as the integral distribution of ${\rm d}N/{\rm 
d}t(>r)\sim 40 \exp(-2.8r)$ s$^{-1}$ or $\sim 4 \int_r^\infty {\rm d} r'\exp(-4 
\ln^2 r')$ s$^{-1}$ (using the net exposure of 8320 s, H. Negoro, private 
communication).

For our hard-state flares, we have found the increase of the count rate in 32 ms 
bins (which is significantly lower than the actual increase seen with ms 
resolution) with respect to the preceding 10 s is by a factor of 6.5 for the 
flares 1, 3 and 9 (and slightly less for the others). Then, the rate of events 
stronger or equal than that is $4\times 10^{-7}$ s$^{-1}$, $1.7\times 10^{-6}$ 
s$^{-1}$ for the two above distributions, respectively. Given our hard-state 
exposure of 1.9 Ms, the former and the latter distribution predicts 0.7, 3.3 
events, respectively. Thus, the flares in the hard state can represent the 
extreme end of the shot distribution, with its log-normal form (proposed 
by Negoro \& Mineshige 2002) preferrred by our data. 

Although shots in the soft state have been studied by F99, no distribution of 
their amplitudes is available. If we still apply the formula for the log-normal 
distribution in the hard state to the obtained relative increase of the 32-ms 
count rate of 9.2 for the flare 13, we obtain one predicted event per $1.3\times 
10^8$ s, i.e., a much longer time interval than our soft-state exposure. Thus, 
the distribution of flare amplitudes in the soft state has most likely a 
different form. 

\section*{ACKNOWLEDGMENTS}

This research has been supported by grants 5P03D00821, 2P03C00619p1,2, 
PBZ-KBN-054/P03/2001 (KBN) and the Foundation for Polish Science. We thank B. 
Stern for the BATSE data, K. Jahoda for help with \xte\/ data processing, and 
the referee and H. Negoro for valuable comments.

%\bsp


\begin{thebibliography}{}

\bibitem{}
Baganoff F. K. et al., 2001, Nat, 413, 45

\bibitem{b99}
Beloborodov A. M., 1999, in Poutanen J. \& Svensson R., eds., High
Energy Processes in Accreting Black Holes. ASP Conf.\ Ser.\ Vol.\ 161,
San Francisco, ASP, p.\ 295

\bibitem{b97}
Boller T., Brandt W.N., Fabian A.C., Fink H.H., 1997, MNRAS, 289, 393

\bibitem{b77}
Boldt, E., 1977, Ann.\ N. Y. Ac.\ Sci., 302, 329

\bibitem{bow65} 
Bowyer S., Byram E.T., Chubb T.A., Friedman M., 1965, Sci, 147, 394

\bibitem{}
Brandt W. N., Boller T., Fabian A. C., Ruszkowski M., 1999, MNRAS, 303, L53

\bibitem{}
Chaput, C. et al., 2000, ApJ, 541, 1026 

\bibitem{}
Churazov E., Gilfanov M., Revnivtsev M., 2001, MNRAS, 321, 759

\bibitem{coppi99}
Coppi P. S., 1999, in Poutanen J. \& Svensson R., eds., High
Energy Processes in Accreting Black Holes, ASP Conf.\ Ser.\ Vol.\ 161,
San Francisco, ASP, p.\ 375

\bibitem{d01}
Di Salvo T., Done C., \.Zycki P. T., Burderi L., Robba N. R., 2001, ApJ, 547,
1024

%\bibitem{esin97}
%Esin A. A., McClintock J. E., Narayan R., 1997, ApJ, 489, 865

\bibitem{}
Feng Y.~X., Li T.~P., Chen L., 1999, ApJ,  514, 373 (F99)

\bibitem{}
Frontera F. et al., 2001, ApJ, 546, 1027

\bibitem{}
Gierli\'nski M., Done C., 2003, MNRAS, in press %(astro-ph/0212384)

\bibitem{}
Gierli\'nski M., Zdziarski A. A., Done C., Johnson W.  N., Ebisawa
K., Ueda Y., Haardt F., Phlips B. F., 1997, MNRAS, 288, 958

\bibitem{}
Gierli\'nski M., Zdziarski A. A., Poutanen J., Coppi P. S.,
Ebisawa K., Johnson W.  N., 1999, MNRAS, 309, 496

\bibitem{}
Golenetskii S., Aptekar R., Frederiks D., Mazets E., Palshin V., Hurley K., Cline T.,  Stern
B., 2003, ApJ, submitted

\bibitem{}
Lin D., Smith I.A., B\"ottcher M., Liang E.P., 2000, ApJ, 531, 963

\bibitem[\protect\citename{Lochner} 1991]{1991ApJ...376..295L} Lochner 
J.C., Swank J.H., Szymkowiak A.E., 1991, ApJ,  376, 295 

\bibitem[\protect\citename{Machida} 2003]{2003ApJ...585..429M} Machida M., 
Matsumoto R., 2003, ApJ,  585, 429 

\bibitem{mz95}
Magdziarz P., Zdziarski A. A., 1995, MNRAS, 273, 837

\bibitem{}
McConnell M. L. et al., 2002, ApJ, 572, 984

\bibitem{}
Meekins J.~F., Wood K.~S., Hedler R.~L., Byram E.~T., Yentis D.~J., Chubb 
T.~A., Friedman H., 1984, ApJ, 278, 288 

\bibitem[\protect\citename{Miyamoto} 1991]{1991ApJ...383..784M} Miyamoto 
S., Kimura K., Kitamoto S., Dotani T., Ebisawa K., 1991, ApJ,  383, 784 

\bibitem{}
Negoro H., Mineshige S., 2002, PASJ, 54, L69 

\bibitem{}
Negoro H., Miyamoto S., Kitamoto S., 1994, ApJ, 423, L127

\bibitem{}
Negoro H., Kitamoto S., Takeuchi M., Mineshige S., 1995, ApJ, 452, L49

\bibitem[\protect\citename{Pottschmidt} 1998]{1998A&A...334..201P} 
Pottschmidt K., Koenig M., Wilms J., Staubert R., 1998, A\&A,  334, 201 

\bibitem{}
Pounds K. A., Done C., Osborne J.  P., 1995, MNRAS, 277, L5

\bibitem{}
Press W. H., Schechter P., 1974, ApJ, 193, 437

\bibitem{}
Proga D., Begelman M. C., 2003, ApJ, in press

\bibitem{}
Revnivtsev M., Gilfanov M., Churazov E., 2000, A\&A, 363, 1013

\bibitem{}
Rothschild R. E., Boldt E. A., Holt S. S., Serlemitsos P. J., 1974, ApJ, 189, L13

\bibitem{}
Rothschild R.~E., Boldt E.~A., Holt S.~S., Serlemitsos P.~J., 1977, ApJ,  
213, 818 

\bibitem{}
Stern B.E., Beloborodov A.M., Poutanen J., 2001, ApJ, 555, 829

\bibitem{}
Sunyaev R., Revnivtsev M., 2000, A\&A, 358, 617 

\bibitem[\protect\citename{Uttley} 2001]{2001MNRAS.323L..26U} Uttley P., 
McHardy I.M., 2001, MNRAS,  323, L26 

\bibitem{}
Weisskopf M.C., Sutherland P.G., 1978, ApJ, 221, 228 

\bibitem{}
Zdziarski A. A., Poutanen J., Paciesas W.S., Wen L., 2002, ApJ, 578, 357 (Z02)

\label{lastpage}

\end{thebibliography}
\end{document}